\documentstyle[preprint,aps]{revtex}
\tightenlines
\begin{document}
\draft
\title{Mesons as $\bar{q}-q$ Bound States from Euclidean 2-Point Correlators
in the Bethe-Salpeter Approach}
\author{T. Meissner and L.S. Kisslinger}
\address{Department of Physics, Carnegie Mellon University, 
Pittsburgh, PA 15213, U.S.A.}

\date{\today} 

\maketitle
 
\begin{abstract}
We investigate the 2-point correlation function 
for the vector current.
The gluons provide dressings for both the quark self energy
as well as the vector vertex
function, which are described consistently by the rainbow Dyson-Schwinger equation and the
inhomogeneous ladder Bethe-Salpeter equation.
The form of the gluon propagator at low momenta is modeled by a 2-parameter ansatz fitting 
the weak pion decay constant $f_\pi$.
The quarks are confined in the sense that 
the quark propagator does not have a pole at timelike 
momenta.
We determine the ground state mass $m_0$ in the vector channel 
from the Euclidean time Fourier transform of the correlator,
which falls off 
as $e^{-m_0 T}$ at large $T$.
$m_0$ lies around
$590 \mbox{MeV}$ and is almost independent of the model form for the gluon propagator.
This method allows us to stay in Euclidean space and to avoid analytic continuation 
of the quark or gluon propagators into the timelike region.
\end{abstract}


\section{Introduction}
The purpose of the present work is the determination of the ground state mass
of the vector $\bar{q}q$ bound state within the Dyson-Schwinger (DS) and
Bethe-Salpeter (BS) approach \cite{iz,rw,tandy}. 
The starting point is the
2-point correlation function 
\begin{equation}
\Pi_{{\mu}{\nu}} (q) = \int d^4x e^{iq \cdot x}
\langle  {\cal{T}} J_{\mu} (x) J_{\nu} (0) \rangle 
\label{corr4}
\end{equation} 
for the vector quark currents $J_\mu (x) = \bar{q} (x) \gamma_\mu q (x)$, with 
$q(x)$ a quark field. 
Within the
approximations used in the present paper, which will be described below,
$\Pi_{{\mu}{\nu}}$ is determined by the 
dressed quark propagator 
\begin{math} 
\langle {\cal{T}} \bar{q} (x) q(0) \rangle 
\end{math}  
and the 
vertex 3 point function
\begin{math}
\langle  {\cal{T}} 
{\bar q} (x) J_{\mu} (y) q (0) \rangle 
\end{math}.
The objective of this work is
to study the large-time behavior of the 2-point correlator in Euclidean space 
in order to determine the
mass of the lowest vector meson, as in lattice gauge calculations \cite{cghn}
and instanton models \cite{shuryak}. 
QCD sum rules
also consider the Euclidean 2-point correlator, but attempt to determine the
ground state mass by using the Borel transform to isolate the lowest mass
pole \cite{svz}.
Our method should be seen in contrast to an earlier approach, which solves the BS 
equation on shell \cite{frank}. 

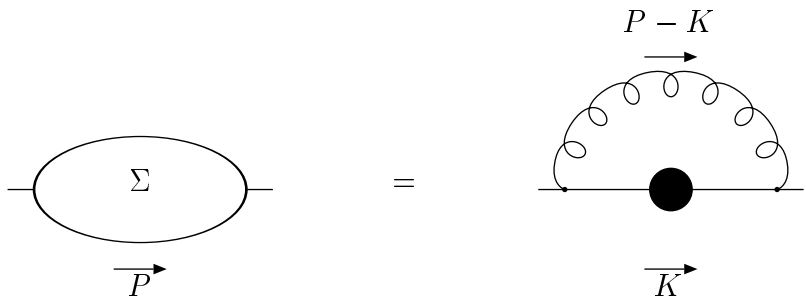
\begin{figure}
\caption{Rainbow Dyson-Schwinger equation for the quark self energy $\Sigma$.}
\label{fig1}
\end{figure}

We now briefly review the DS-BS approach used here.
The nonperturbative nature of QCD at low and intermediate momentum transfers
makes the use of effective models a natural and necessary tool.
During the past few years strong progress has been made in the framework
of the Dyson-Schwinger (DS) approach \cite{iz,rw,tandy},
which is based on a coupled set of relations between dressed quark,
gluon and ghost propagators and vertex functions.
In order to handle this system it is of course necessary to make certain 
simplifications and truncations.
One approach, which is commonly referred to as {\it rainbow} approximation, 
uses an undressed quark-gluon vertex and solves the Dyson-Schwinger (DSE) for 
the dressed quark
propagator $G(p)$ with a given dressed gluon propagator $D_{\mu\nu}^{ab} (p)$ 
as input (c.f. Fig.\ref{fig1}):
\begin{equation}
\Sigma(p)  =  \int \frac{d^4q}{(2\pi)^4}  {g_s}^2 D_{\mu\nu} ^{ab} (p-q) 
\gamma_\mu \frac{\lambda_c ^a}{2} G(q)\gamma_\nu \frac{\lambda_c ^b}{2}
\quad .
\label{sd}
\end{equation}
The quark propagator $G$ and the quark self energy $\Sigma$ are related by:
\begin{equation}
G (p)^{-1}  =  i\gamma \cdot p + \Sigma(p) \quad .
\label{eq-sig}
\end{equation}
The form of the gluon propagator $D_{\mu\nu}^{ab} (p)$ 
for small momenta is basically unknown from QCD and therefore has
to be modeled while following certain requirements which will be discussed 
below. In case of this special truncation it is possible to formulate an
effective field theory using functional integral techniques, which is known
as the Global Color Model (GCM)\cite{rw,tandy,cahill1}.
This has the striking advantage that in the chiral limit (zero current mass 
for the light quarks, i.e. $M_u = M_d = M_s =0$) 
one is able to establish an easy connection between
the dressed quark self energy leading to dynamical chiral symmetry breaking on
one side and the existence of massless Goldstone bosons as pseudoscalar
$\bar{q}q$ bound states, on the other side.
As a consequence it is possible to perform a systematic chiral low energy 
expansion \cite{fm1} in terms of composite Goldstone boson field and
therefore make a connection between
the phenomenological chiral hadronic theory (Chiral Perturbation Theory) \cite
{chiral} and an effective quark based theory of QCD, the GCM. \\
Furthermore one is also able to study the coupling of composite $\bar{q}q$ 
systems to an external 
electromagnetic gauge field while guaranteeing local $U(1)$ gauge invariance 
which is manifest in the corresponding Ward-Takahashi identity
\cite{ft,frank}. It turns out that the chiral invariance
and electromagnetic current conservation,
which are both essential for studying low energy hadronic phenomena,
are very difficult to maintain if one tries to go beyond the rainbow 
approximation\cite{smekal} and are lost in general if 
model ans\"atze for the quark-gluon vertex are used.
 
Though the rainbow approximation and the GCM naturally 
violate local color $SU(3)_c$ gauge invariance and renormalizability,
they provide a very successful description of various nonperturbative aspects 
of strong interaction physics and the QCD vacuum as well as hadronic
phenomena at low energies.
These include for instance quark confinement \cite{rw}, 
QCD vacuum condensates \cite{cahill1,meissner,km,maris2}, 
$U_A (1)$ breaking and the $\eta-\eta^\prime$ splitting \cite{fm2},
low energy chiral dynamics of Goldstone bosons ($\pi,K,\eta$) 
\cite{roberts,fm1,gunner,bqrtt,maris1,maris2},
meson form factors \cite{formfactor}, heavy-light mesons \cite{heavy},
systems at finite temperature \cite{temperature} or
soliton \cite{soliton} and Fadeev \cite{cahill2}  descriptions of the nucleon.

The method is more involved when applied to higher mass states 
instead of Goldstone bosons, e.g. $\rho$, $\omega$, $\sigma$ or $A$ mesons.
A mesonic $\bar{q}q$ bound state with mass $m$ in a channel with
spin-flavor index $\theta$ is determined in the ladder approximation as
solution of the {\it homogeneous BSE} \cite{frank}: 
\begin{equation}  
\Omega_\theta (P,q^2 = - m^2) = 
\int \frac{d^4 K}{(2\pi)^4} {g_s}^2 D_{\mu\nu} ^{ab} (P-K) 
\gamma_\mu \frac{\lambda_c ^a}{2} G(K_+) 
\Omega_\theta (K,q^2 = - m^2) G(K_-) \gamma_\nu \frac{\lambda_c ^b}{2} \quad ,
\label{bs}
\end{equation}
where we have used the notation $K_{\pm} = K \pm \frac{q}{2}$. 
The $\Omega_\theta (P,q)$ is related to the Bethe-Salpeter amplitude 
$\chi_\theta (P,q)$ as:
\begin{equation}
\chi_\theta (P,q^2 = - m^2) = G(P_+) 
\Omega_\theta (P,q^2 = - m^2) G(P_-) \quad . 
\label{bsa}
\end{equation}
 
As mentioned above in the pseudoscalar channel the existence of a massless 
state or in other words
a solution of (\ref{bs}) at $q^2 =0$ is automatically guaranteed in the chiral
limit if the rainbow DSE (\ref{sd}) is satisfied. 
For a small finite current quark mass $M_0$ a perturbative expansion in $M_0$ and $q^2$
can be performed.
This is however not possible for larger meson masses
and therefore an explicit numerical solution
of the corresponding homogeneous BSE (\ref{bs}) is required.
The trouble hereby is that the meson momentum is defined in the timelike 
region, $q^2 = - m^2$, which in turn requires
the dressed quark propagator $G(p)$ in the timelike region.
However $G(p)$ is a priori only determined at spacelike $p$  from the DSE 
(\ref{sd}), because the model gluon propagator $D(p)$ is defined only for
spacelike Euclidean momenta.
It is therefore necessary to perform an analytic continuation of the dressed
quark propagator $G(p)$ from
Euclidean space into Minkowski space $p_4 \to i p_0$ \cite{frank}. 
This is a very dangerous and unsafe procedure because it requires in fact
the knowledge of the singularity structure of $G(p)$ in the whole complex 
plane, which we do not have.
In Ref. \cite{bfm} an alternative method has been proposed which 
stays completely in Euclidean space
avoiding analytic continuations of any of the dressed propagators.
It is the analog to what is used in instanton models \cite{shuryak} and 
lattice QCD calculations \cite{cghn}
where working in Euclidean space is essential.
 
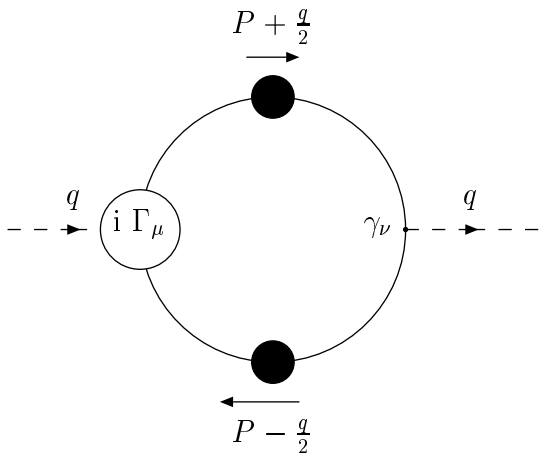
\begin{figure}
\caption{2-point correlator $\Pi_{\nu\mu}$ in the vector channel.}
\label{fig2}
\end{figure}

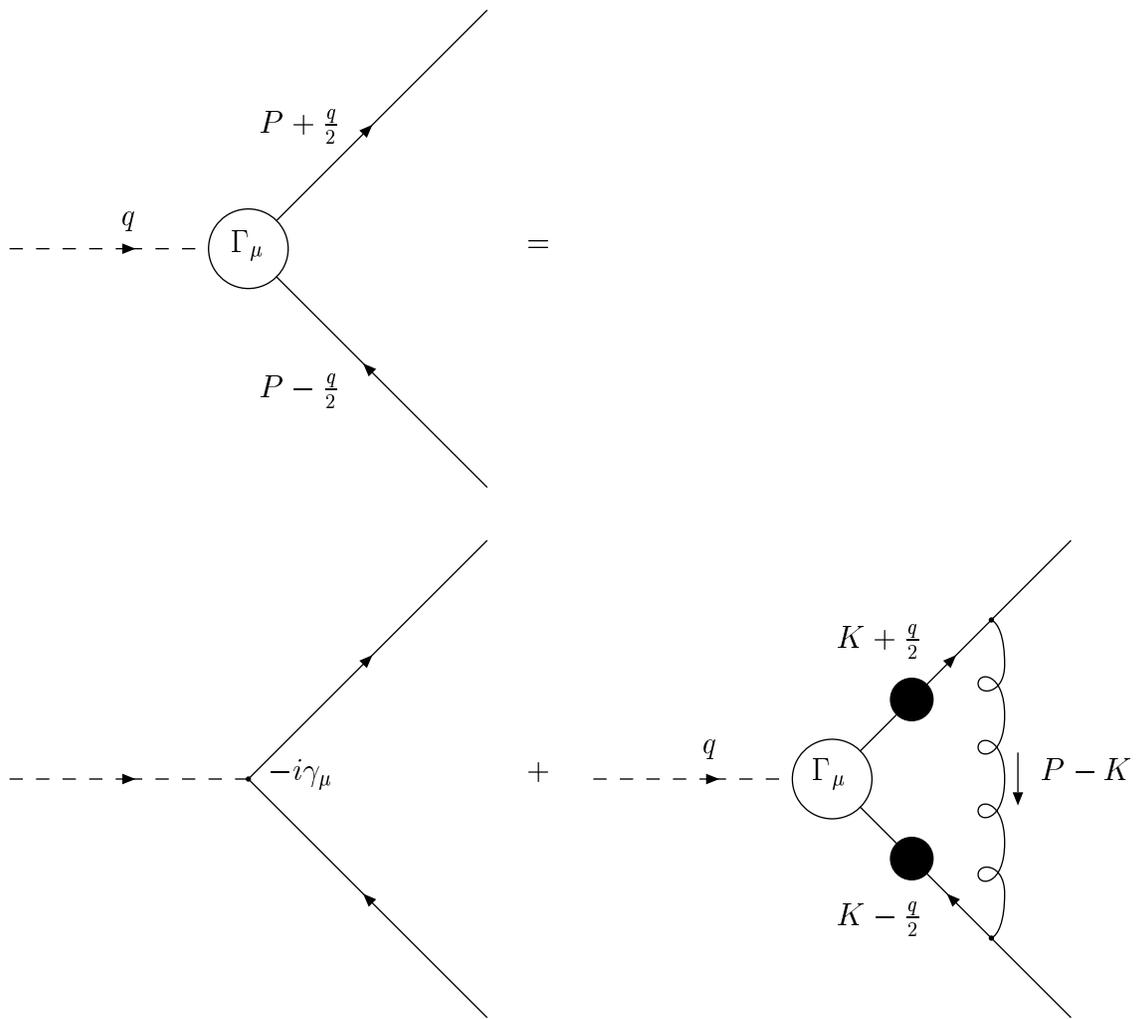
\begin{figure}
\caption{Inhomogeneous ladder Bethe-Salpeter equation 
for the Vector Vertex Function $\Gamma_\mu$.}
\label{fig3}
\end{figure}

The 2-point correlation function 
\begin{equation}
\Pi_{{\theta_1}{\theta_2}} (q) = \int d^4x e^{i q \cdot x}
\langle  {\cal{T}} J_{\theta_1} (x)  J_{\theta_2} (0) \rangle 
\label{corr1}
\end{equation} 
with quark currents $J_\theta = \bar{q} T_\theta q$, where 
$\theta$ stands for the spin-flavor of the operator $T_\theta$,
is determined as a quark loop containing the dressed vertex function
$\Gamma_\theta (P,q)$ (c.f. Fig.\ref{fig2} for $T_\theta = \gamma_\mu$).
The $\Gamma_\theta (P,q)$ itself is given
as the solution of the {\it inhomogeneous} ladder BSE (c.f. Fig.\ref{fig3}):
\begin{equation}
\Gamma_\theta (P,q) = -i T_\theta +  
\int \frac{d^4 K}{(2\pi)^4} {g_s}^2 D_{\mu\nu} ^{ab} (P-K) \gamma_\mu 
\frac{\lambda_c ^a}{2} G(K_+) \Gamma_\theta (K,q) G(K_-) \gamma_\nu  
\frac{\lambda_c ^b}{2} 
\label{ibse}
\end{equation}
with the inhomogeneity $-i T_\theta$. This form can also be written as
\begin{equation}
\Gamma_\theta (P,q) = -i T_\theta +  \Gamma_\theta (P,q)^{NP},
\label{ibse2}
\end{equation}
where T$_\theta$ is the bare vertex and
$\Gamma_\theta (P,q)^{NP}$ is the nonperturbative dressed vertex.
The nonperturbative parts of the vertex functions
can be constrained by the forms of the nonlocal condensates \cite{jk},
which also has been used in our recent work \cite{km}. Moreover, it has
been shown that this formalism can be used for the nonperturbative part of
the vector vertex function used at low momentum transfer for nuclear magnetic
dipole moments\cite{lsk}. This gives possible constraints for the dressed
vertex function treated in the present work, which we discuss below.
 
The momentum $q$ is in general off shell and can be either space or timelike.
Close to the mass shell $q^2 \approx - m^2$ the $\Gamma_\theta(P,q)$ and
the solution of the homogeneous BSE (\ref{bs}) $\Omega_\theta(P,q^2 = - m^2)$
are related by
\begin{equation}
\Gamma_\theta(P,q^2 \approx - m^2) \approx \frac{\Omega_\theta(P,q^2 = - m^2)}
{q^2 + m^2}
\quad .
\label{rel}
\end{equation}
In our study the $q^2$ will be always spacelike. 
Other than the dressed quark or gluon propagator the correlator 
$\Pi_{\theta_1 \theta_2} (q^2)$ has a  
known Kall\'{e}n-Lehmann spectral representation \cite{bd}: Its poles are at 
$q^2 = - {m_I}^2 ,\, I =0,1,\dots $, where the $m_I$ are the meson resonance 
masses and the pole
residua are basically given by the couplings $f_I$ between the interpolating 
current $J_\theta$ and the on-shell states $\vert I \rangle $.
Otherwise $ \Pi_{\theta_1 \theta_2} (q^2)$ is holomorph in the complex plane 
and falls off sufficiently fast for $\vert q^2 \vert \to \infty$, so that
Cauchy's theorem can be applied.
A convenient method to filter out the lowest state with mass $m_0$ is to take 
the the one dimensional
Fourier transform of (\ref{corr1}) with respect to the time component of the 
Euclidean momentum $q_4$
\begin{equation}
\Pi_{\theta_1 \theta_2}^* (T) = \int \frac{d q_4}{2\pi} e^{iq_4 T} \Pi_{\theta
 _1 \theta_2} (q_4,\vec{q}=0) 
\label{ft}   
\end{equation}
which falls off as $e^{-m_0 T}$ for large Euclidean times $T$. 
 
It is obvious that this method will work in reality only if the ground state 
mass $m_0$ is
sufficiently well separated from the mass of the first excited state $m_1$. 
If $m_0$ and $m_1$ are too close it will be numerically
difficult to extract the exponential falloff, because this situation will 
require numerical very large
values for $T$ in order to see the falloff and for those large values the 
signal is likely to be swamped by numerical errors due to highly oscillating
behavior of $e^{iq_4 T}$.
On the other hand it is clear that we have indeed avoided to make any 
assumption about the 
analyticity of the propagators. Instead we are using the analyticity structure
of the 2-point correlator $ \Pi_{\theta_1 \theta_2} (q^2)$ in the complex
plane, which is well under control.
The price which we have to pay is the fact that we have to solve 
the inhomogeneous BSE over the whole range of
spacelike $q^2$ instead of the homogeneous BSE at only the on shell point $q^2
 = - m_0 ^2$.
 
It is the aim of our paper to perform the first analysis of the 2-point 
correlation function
in the vector channel (i.e. $T_\theta = \gamma_\mu$) using this approach and 
to extract the ground state mass $m_0$.
We will work in the chiral limit ($M_u = M_d = M_s =0$), therefore isospin 
effects are not present.  Following Ref. \cite{fm1} we will use a model gluon
propagator $D(q^2)$ with an IR regularized Mandelstam $\frac{1}{q^4}$
singularity \cite{mandelstam,marciano} at $q^2 \approx 0$.
For large $q^2$ an asymptotic free UV tail is added.
The parameters in $D(q^2)$ are fixed in such a way that the weak pion decay 
constant in the chiral limit $f_\pi = 88 \mbox{MeV}$ is reproduced and, as 
it was shown in Ref. \cite{fm1,meissner,km} a 
satisfactorily description of low energy chiral physics and vacuum condensates
can be achieved. We demonstrate that the dressed quarks which are
determined from the rainbow DSE (\ref{sd}) are
confined in the sense that the quark propagator does {\it not} show an 
exponential falloff $\sim e^{-M T}$ for large $T$.
We then demonstrate the existence of a bound state in the vector channel by 
showing the exponential falloff of the correlator $\Pi_{\mu\mu} ^* (T)$ for
large $T$ from which we can determine 
the ground state mass $m_0$.
 
Our paper is organized as follows:
In section 2 we briefly review the 
definition of the dressed quark propagator $G$ and the
vector vertex $\Gamma_\mu$ as well as the 
rainbow DSE for $G$ and the inhomogeneous BSE for
$\Gamma_\mu$ 
and demonstrate the validity of the vector Ward-Takahashi identity. 
We then show in section 3 
how to obtain the expression for the vector 2-point correlator and its
Fourier transform in Euclidean space.
The numerical results for gluon and quark propagators, vertex function and 
correlator are presented, analyzed and discussed in section 4. 
Finally, conclusions are offered in section 5.
 
\section{Dressed Quark Propagator and Vertex Function}
\subsection{Rainbow Dyson-Schwinger and Ladder Bethe Salpeter Equation}
Following Ref. \cite{ft} the inverse of the
dressed quark propagator G(p) has the form
\begin{equation}
G^{-1} (p) \equiv   i \gamma \cdot p [A(p^2) -1] + \Sigma(p) 
\equiv i \gamma \cdot p A(p^2) + B(p^2) 
\label{quarkpropagator}
\end{equation}
in momentum space.
The quark self energy dressing $\Sigma (p)$ comprises the 
concept of {\it constituent} quarks by providing a running mass
$M(p^2)=B(p^2)/A(p^2)$.
It is determined as the solution of the rainbow DSE 
\begin{equation}
\Sigma (p) = \frac{4}{3}{g_s}^{2} \frac{d^{4}q}{(2\pi)^{4}}
D (p-q) \gamma_\nu G (q) \gamma_\nu
\label{sdesigma}
\end{equation}
which is schematically shown in Fig.\ref{fig1}.
For convenience the Feynman type gauge for the gluon propagator  
\begin{equation}
D_{\mu\nu}^{ab} (q) = \delta_{\mu\nu} \delta^{ab} D(q^2)
\label{feynmangauge}
\end{equation}
has been used here, which defines the phenomenological function,
$D(q^2)$, in Eq.(\ref{sdesigma}). 
In the rainbow DS-BS approximation
used in the present work, this function $D(q^2)$ for the dressed gluon
propagator is our main physical input.

In terms of the components $A$ and $B$ Eq.(\ref{sdesigma}) reads:
\begin{mathletters}
\label{sde}
\begin{eqnarray}
\left[ A(p^{2})-1\right] p^{2} &=& {g_s}^{2} 
\frac{8}{3}\int \frac{d^{4}q}{(2\pi 
)^{4}}
D(p-q)\frac{A(q^{2})q\cdot p}{q^{2}A^{2}(q^{2})+B^{2}(q^{2})} 
\label{sde:a} \\
B(p^{2})
&=&
{g_s}^{2}\frac{16}{3}\int \frac{d^{4}q}{(2\pi 
)^{4}}D(p-q)\frac{B(q^{2})}{q^{2}A^{2}(q^{2})+B^{2}(q^{2})} \;  ,
\label{sde:b}
\end{eqnarray}
\end{mathletters}
 
In momentum space the dressed vector vertex is 
determined by the {\it inhomogeneous ladder BSE}
which reads in momentum space 
(\ref{ibse}) 
\begin{equation}
\Gamma_\mu (P,q) = (-i) \gamma_\mu - 
\frac{4}{3} {g_s}^2  \int \frac{d^4 K}{(2\pi)^4} D(P-K) \gamma_\nu
G(K+\frac{q}{2}) \Gamma_\mu (K,q) G(K-\frac{q}{2}) \gamma_\nu
\label{ibsemomentum}
\end{equation}
and is schematically displayed in  Fig.\ref{fig3}. 
 
As has been shown in Ref. \cite{ft} both the rainbow DSE (\ref{sdesigma}) 
and the ladder BSE
(\ref{ibsemomentum}) can be {\it consistently} derived from the action of the 
global color model (GCM) in an external gauge field ${\cal{A}}_\mu (z)$ using 
standard functional integration
techniques. 
The main steps of this derivation are reviewed in Appendix A and B.
 
A crucial consequence is that this 
formalism ensures by construction invariance under local $U(1)$ gauge 
transformations. 
At the level of the dressed vertex $\Gamma_\mu$ this invariance
is reflected in the validity of the {\it Ward-Takahashi identity (WTI)}:
\begin{equation}
q_\mu \Gamma_\mu (P,q)  = 
G^{-1} (P-\frac{q}{2}) - G^{-1} (P+\frac{q}{2}) \, ,
\label{wti}
\end{equation}
which can be easily verified by substituting it into (\ref{ibsemomentum}) and 
using the DSE (\ref{sdesigma}) for the quark self energy $\Sigma$ as well as
the definition (\ref{quarkpropagator}).\\ 
Expanding the r.h.s. of (\ref{wti}) in $q_\mu$ and taking the limit 
$q_\mu \to 0$ leads to the {\it Ward identity (WI)}:
\begin{equation}
\Gamma_\mu (P,0) = - \frac{\partial G^{-1} (P) }{\partial P_\mu } \quad  .
\label{wi}
\end{equation} 
 
\subsection{General Form of the Dressed Vertex $\Gamma_\mu (P,q)$}
Following Ref. \cite{frank} in the Feynman-type gauge (\ref{feynmangauge}) 
for the model gluon propagator the most general form for the vertex function
$\Gamma_\mu (P,q)$ which fulfills (\ref{ibsemomentum}) reads
\begin{equation}
\Gamma_\mu (P,q) = \openone  \Lambda_\mu ^{(1)} (P,q) 
+ i \gamma_\nu  \Lambda_{\nu\mu} ^{(2)} (P,q) 
+ i \gamma_5 \gamma_\nu \epsilon_{\mu\nu\alpha\beta} P_\alpha q_\beta \eta^{-2
} \Lambda^{(3)} (P,q) \, ,
\label{decomp1}
\end{equation}
where for convenience 
an arbitrary mass scale $\eta$ has been introduced on order to render all the 
$\Lambda^{(i)}$ dimensionless.
Furthermore it turns out to be convenient to perform a decomposition into 
longitudinal
and transversal components with respect to the external momentum $q_\mu$:
\begin{mathletters}
\label{decomp2}
\begin{eqnarray}
\Lambda_\mu ^{(1)} (P,q) &=& \frac{q_\mu}{\eta} \frac{P\cdot q}{q^2} \lambda_1
  ^L 
+ \frac{P_\mu ^T}{\eta}  \lambda_1 ^T  
\label{l1} \\
 \Lambda_{\nu\mu} ^{(2)} (P,q) &=& 
\frac{P_\nu q_\mu}{\eta^2} \frac{P\cdot q}{q^2}  \lambda_2 ^L + \frac{P_\nu P_
\mu ^T}{\eta^2}
\lambda_2 ^T \nonumber \\ 
&& - \frac{q_\nu q_\mu}{q^2}   \lambda_3 ^L - \left (\delta_{\nu\mu} - \frac{q
_\nu q_\mu}{q^2} 
\right ) \lambda_3 ^T \nonumber \\ 
&& + \frac{q_\nu P_\mu ^T P\cdot q }{\eta^4} \lambda_4 ^T \\ \label{l2}
 \Lambda^{(3)} (P,q) &=& \lambda_5 ^T  \, .
\label{l3}
\end{eqnarray}
\end{mathletters}
The eight scalar dimensionless coefficients $\lambda_i ^L$ ($i=1,\dots,3$)
and $\lambda_i ^T$
($i=1,\dots,5$) depend on $P^2$, $q^2$ and $C_{Pq} ^2$,
where $C_{Pq} = \frac{P\cdot q}{q^2}$
is the direction cosine between $P$ and $q$.
$P_\mu ^T \equiv P_\mu - \frac{P\cdot q q_\mu}{q^2}$ is the vector transverse 
to $q_\mu$,
i.e. $q_\mu P_\mu ^T =0$.
The advantage of the decomposition (\ref{decomp2}) lies in the fact that the 
longitudinal
components  $\lambda_i ^L$ ($i=1,\dots,3$) are determined automatically from 
the quark propagator $G$ by means of the WTI (\ref{wti})
\begin{mathletters}
\label{llong}
\begin{eqnarray}
\frac{P\cdot q}{\eta}  \lambda_1 ^L &=&
B(P_- ^2) -B(P_+ ^2) \label{llong1}  \\
\frac{P\cdot q}{\eta^2}  \lambda_2 ^L &=&
A(P_- ^2) -A(P_+ ^2) \label{llong2}  \\
2 \lambda_3 ^L &=& A(P_- ^2) + A(P_+ ^2)  \label{llong3}  \,  ,   
\end{eqnarray}
\end{mathletters}
where $P_{\pm} \equiv P \pm \frac{q}{2}$. 
 
This leaves only the 5 independent transversal components
$\lambda_i ^T$ ($i=1,\dots,5$) to be determined as solutions of the 
inhomogeneous BSE (\ref{ibsemomentum}).
In order to do this one has to project out the single $\lambda_i ^T$ for each 
$i$ which can be done by multiplying (\ref{decomp1}) and (\ref{decomp2}) 
with appropriate Dirac matrices and taking traces.
Performing the same operations on the r.h.s of (\ref{ibsemomentum}) leads 
finally to a set of 5 coupled linear inhomogeneous integral equations for the 
$\lambda_i ^T$ ($i=1,\dots,5$) \cite{frank}:
\begin{eqnarray}
{\cal{X}}_i &=& 
{\cal{N}}_{ij} \lambda_j ^T (P^2,q^2,C_{Pq} ^2 ) \, + \, \nonumber \\
&+& 
\int_0^\infty dK K^3 \int_{-1}^{+1} d C_{Kq} {\cal{M}}_{ij} (q^2,P^2,C_{Pq} ^2
,K^2,C_{Kq} ^2)   
 \lambda_j ^T (K^2,q^2,C_{Kq} ^2 ) \, .
\label{integraleq}
\end{eqnarray}
The indices $i$ and $j$ are both running from $1$ to $5$.
The vector ${\cal{X}}_i$ and the nonzero components of the matrices
${\cal{N}}_{ij}$ and ${\cal{M}}_{ij}$ have been derived in 
Ref.\cite{frank}. 
For completeness and because we need them later for the determination of the correlator they are  
also listed in Appendix C.

\subsection{Asymptotic Behavior}
\label{sec:asy}
For large values of the quark momentum $P^2 \to \infty$ or for large values 
of the external momentum
$q^2 \to \infty$ the propagators reach their asymptotic free forms
and therefore the kernel in (\ref{ibsemomentum})
vanishes, which implies that also $\Gamma_\mu$ approaches the naked vertex.
This means:
\begin{equation}
\Gamma_\mu (P,q) \vert_{P^2\to\infty} = (-i) \gamma_\mu
\label{asy1}
\end{equation}
or
\begin{mathletters}
\label{asy2}
\begin{eqnarray}
\lambda_i ^L (P^2\to\infty,q^2,C_{PQ} ^2) &=& 0 \; , \; i=1,2  \label{asy2a} 
\\
\lambda_i ^L (P^2\to\infty,q^2,C_{PQ} ^2) &=& 1 \; , \; i=3   \label{asy2b} \\
\lambda_i ^T (P^2\to\infty,q^2,C_{PQ} ^2) &=& 0 \; , \; i=1,2,4,5 \label{asy2c
} \\
\lambda_i ^T (P^2\to\infty,q^2,C_{PQ} ^2) &=& 1 \; , \; i=3 \;  ,  \label{asy2
d}
\end{eqnarray}
\end{mathletters}
and 
\begin{equation}
\Gamma_\mu (P,q) \vert_{q^2\to\infty} = (-i) \gamma_\mu
\label{asy3}
\end{equation}
or
\begin{mathletters}
\label{asy4}
\begin{eqnarray}
\lambda_i ^L (P^2,q^2\to\infty,C_{PQ} ^2) &=& 0 \; , \; i=1,2  \label{asy4a} 
\\
\lambda_i ^L (P^2,q^2\to\infty,C_{PQ} ^2) &=& 1 \; , \; i=3   \label{asy4b} \\
\lambda_i ^T (P^2,q^2\to\infty,C_{PQ} ^2) &=& 0 \; , \; i=1,2,4,5 \label{asy4c
} \\
\lambda_i ^T (P^2,q^2\to\infty,C_{PQ} ^2) &=& 1 \; , \; i=3   \label{asy4d}
\end{eqnarray}
\end{mathletters}
together with $A(P^2\to\infty)=1$ and $B(P^2\to\infty)=0$.
 
\subsection{Constraints on Vertex Function}
The nonperturbative part of the vector vertex function from Eqs.
(\ref{ibse},\ref{ibse2}) can be constrained by the
work using QCD sum rules with nonlocal condensates.  
$\Gamma_\mu (y_1,y_2;z)^{NP}$, defined as the second term in 
Eq.(\ref{ibsecoordinate}) 
has also been shown to be given by the four-quark nonlocal 
condensate\cite{jk,lsk}
\begin{equation}
 \Gamma_\mu (y_1,y_2;z)^{NP}  =  
 \langle 0 \vert :q(y_1) \bar{q} (z) \gamma_\mu q(z) \bar{q} (y_2) : \vert 0 \rangle \, .
\label{3p2p}
\end{equation}
Defining the nonlocal quark condensate by
\begin{equation}
 \langle 0 \vert : \bar{q} (0) q (y) : \vert 0 \rangle
   \equiv  g (y^2) \, \langle 0 \vert : \bar{q} (0) q(0) : \vert 0
\rangle \, ,
\label{nlc}
\end{equation}
the function $g(y^2)$ gives the space-time structure of the nonlocal
condensates. 
In particular, the vertex function $\Gamma_\mu (0,0;z)^{NP}$,
needed in the three-point determination of the vector vacuum susceptibility
is given by Eqs.(\ref{3p2p},\ref{nlc}) with the approximation of
vacuum factorization as
\begin{equation}
 \Gamma_\mu (0,0;z)^{NP}  =  (- i \gamma_\mu)
\left ( \frac{\langle  \bar{q} q \rangle}{12} \right )^2
\int d^4y g (y^2)  g ((z-y)^2 )   \,  .
\label{gammaz}
\end{equation}
The vacuum factorization approximation has been found to be satisfactory for
deriving vacuum susceptibilities \cite{jk,lsk}. 
Using the forms of $g(y^2)$ such as those studied in Refs.\cite{km,jk,lsk}
one obtains constraints on the dressed vertex functions and thus further
constraints on the gluon propagator. This will be the subject of future work.

\section{Vector Correlator}
 
\subsection{Definition and Properties}
We are now ready to study the main object of our investigation which is the 
correlator
$\Pi_{\mu_1\mu_2}$ of 2 vector currents $j_\mu (x) = \bar{q} (x) \gamma_\mu 
q(x)$.
In momentum space it is defined as:
\begin{equation}
\Pi_{\mu_1\mu_2} (q) 
\equiv  
\int d^4 x e^{iqx} \left \langle  {\cal{T}} j_{\mu_1} (x) j_{\mu_2} (0) 
\right \rangle \,  .
\label{pidef}
\end{equation}
 
It can be expressed in terms of the dressed quark propagator $G$ and the 
dressed vertex function
$\Gamma_\mu (P,q)$ as
\begin{equation}
\Pi_{\nu\mu} (q) = 
i \int \frac{d^4 P}{(2\pi)^4} \mbox{Tr} \left [
\Gamma_\mu (P,q) G(P_+) \gamma_\nu G(P_-) \right  ]  \, ,
\label{corr3}
\end{equation}
which is schematically displayed in Fig.\ref{fig2}.
The trace $\mbox{Tr}$ goes over Dirac and color indices.
A derivation of (\ref{corr3}) from the GCM is given in Appendix D.
It should be stressed that (\ref{corr3}) is a mean field result.
 
The WTI (\ref{wti}) for the dressed vertex $\Gamma_\mu$ ensures that the 
correlator $\Pi_{\nu\mu} (q)$ is {\it transversal}
\begin{equation}
q_\mu \Pi_{\nu\mu} (q) = 0
\label{trans1}
\end{equation}
or
\begin{equation}
\Pi_{\nu\mu} (q) = \left ( \delta_{\mu\nu} - \frac{q_\mu q_\nu}{q^2} \right ) 
\Pi (q^2) \; .
\label{trans2}
\end{equation}
This can be easily verified by putting (\ref{wti}) into (\ref{corr3}) and 
reflects
again manifest invariance of our approach under $U(1)$ gauge transformation 
and therefore conservation of the vector current $j_\mu (x)$.  
Because of (\ref{trans2}) it is sufficient to consider the contraction $\Pi_
{\mu\mu} (q^2)$.

\subsection{Spectral Representation and Bound States}  
The Kall\'{e}n-Lehmann spectral representation \cite{bd} for $\Pi_{\mu\mu} (q^
2)$ is obtained by saturating the correlator 
(\ref{pidef}) with a complete set of on-shell mesonic bound states 
$\vert I(m_I,k_I,\lambda_I) \rangle$ ($I=0,1,2,\dots$) 
with mass $m_I$, 4-momentum $P_I$ ($P_I ^2 = - m_I ^2$ in Minkowski space) and
polarization $\lambda_I$:
\begin{equation}
\Pi_{\mu\mu} (q^2) = \int_0^\infty d s {\tilde{\rho}}_{\mu\mu} (s) \frac{1}{s-
q^2 + i \epsilon}
\label{spectral1}
\end{equation}
where
\begin{eqnarray}
\rho_{\mu\mu} (q) &=&  {\tilde{\rho}}_{\mu\mu} (q^2) \theta(q_0)  
\nonumber \\
&=&
(2\pi)^3 \sum_I \delta^{(4)} (P_I - q) 
\langle  0 \vert j_\mu (0) \vert I(m_I,P_I,\lambda_I) \rangle
\langle  I(m_I,P_I,\lambda_I) \vert j_\mu (0) \vert 0 \rangle \; .
\label{spectral2}
\end{eqnarray}
 
The coupling between the on-shell state $\vert I \rangle$ and the 
interpolating quark current $j_\mu$ is denoted by $f_I$ and defined by
\begin{equation}
\langle  0 
\vert
j_\mu (x) \vert  I(m_I,P_I,\lambda_I) \rangle = f_I e^{ikx} \epsilon_\mu(P_I,
\lambda_I) 
\frac{1}{\sqrt{(2\pi)^3 (2{\omega_{P}}_I)}} \; ,
\label{fidef}
\end{equation}
where $ \epsilon_\mu(P_I,\lambda_I)$ denotes the polarization four vector and 
$({\omega_{P}}_I)^2 = m_I ^2 + \vec{k}_I ^2$.
Inserting (\ref{fidef}) in (\ref{spectral2}) leads in Euclidean space to
\begin{equation}
\Pi_{\mu\mu} (q^2) = 3 \, \sum_I  \frac{f_I ^2}{q_I ^2 + m_I ^2} \; .
\label{spectral3}
\end{equation}
 
In order to filter out the contribution from the ground state
$\vert I=0\rangle$ we take the Fourier transform of (\ref{spectral3}) with
respect to the Euclidean time $T$
\begin{equation}
{\Pi_{\mu\mu}}^* (T) = \int_{-\infty}^{+\infty} \frac{dq_4}{2\pi}
e^{iq_4 T}
\Pi_{\mu\mu}(q_4,\vec{q}=0) =
\frac{3}{2} \sum_I \frac{f_I ^2}{m_I} e^{-m_I T} \quad ,
\label{pit}
\end{equation}
and consider the limit $T\to\infty$
\begin{equation}
{\Pi_{\mu\mu}}^* (T\to\infty) \to \frac{3}{2} \frac{f_0 ^2}{m_0} e^{-m_0 T}
\;.
\label{ground}
\end{equation}     
 
\subsection{Determination of $\Pi_{\mu\mu}$ in Terms of $G$ and $\Gamma_\mu$}
The final task which remains in order to evaluate $\Pi_{\mu\mu}$ is to write 
it in terms of the quark self energy functions $A$ and $B$
(\ref{quarkpropagator}) and vertex functions
$\lambda_i ^T$ (\ref{decomp1},\ref{decomp2}).
This is done straightforwardly by inserting 
the definitions 
(\ref{quarkpropagator},\ref{decomp1},\ref{decomp2}) 
into (\ref{corr3}) and carrying out the matrix trace.
The result is
\begin{eqnarray}
\Pi_{\mu\mu} (q^2) = 
(- N_c) \frac{1}{\eta^2\pi^3}  \int_0^\infty dP P^3 \int_{-1}^{+1} dC_{Pq} x 
\cdot
\Bigl \{&-& 
\frac{P^2}{\eta^2} x^2 V \lambda_1 ^T (P^2,q^2,C_{Pq} ^2)  \nonumber \\
&+& 
\frac{P^2}{\eta^2} x^2 \left ( F_1 - 2 \frac{P^2}{\eta^2} T \right )  
\lambda_2 ^T (P^2,q^2,C_{Pq} ^2)  \nonumber \\
&-&
\left ( 3 F_1 - 2 \frac{P^2}{\eta^2} x^2 T \right )  
\lambda_3 ^T (P^2,q^2,C_{Pq} ^2)  \nonumber \\
&-&
2 \frac{P^4 q^2}{\eta^6} x^2 C_{Pq} ^2 T 
\lambda_4 ^T (P^2,q^2,C_{Pq} ^2)  \nonumber \\
&-&
2 \frac{P^2 q^2}{\eta^4} x^2  T 
\lambda_5 ^T (P^2,q^2,C_{Pq} ^2)  \; \Bigr \}  \; ,
\label{picomp}
\end{eqnarray}
where the abbreviations from Appendix C  
have been used.
 
At this point we have to say how to handle the UV divergences in $\Pi_{\mu\mu}
(q^2)$. For large quark loop momenta $P$ the quark propagator and vertex
functions reach their asymptotic forms specified in section \ref{sec:asy} 
and the loop in Fig.\ref{fig2} is the free loop which is logarithmically 
divergent. In perturbation theory it can be renormalized in the standard way
using dimensional  regularization \cite{narison}.
In our case it is actually not necessary to evaluate $\Pi_{\mu\mu} (q^2)$, 
because we need only its Fourier transform ${\Pi_{\mu\mu}}^* (T)$, which is 
UV finite, as any correlator in coordinate space is UV finite.
The reason for that is that the quark propagator, which falls off like
$\frac{1}{P}$ in momentum space, will fall off like $e^{-T P}$ after Fourier 
transform, and therefore the momentum integral 
$\int_0^{\infty} dPP^3$ in (\ref{picomp}) is UV finite.
When evaluating ${\Pi_{\mu\mu}}^* (T)$ from (\ref{picomp})
it is therefore convenient to perform the Fourier transform
$\int_{-\infty}^{+\infty} \frac{dq_4}{2\pi} e^{iq_4T}$ {\it before} 
the momentum integral $\int_0^{\infty} dPP^3$,
which avoids a conceptual and technical intricate renormalization procedure.

\section{Results and Discussion}
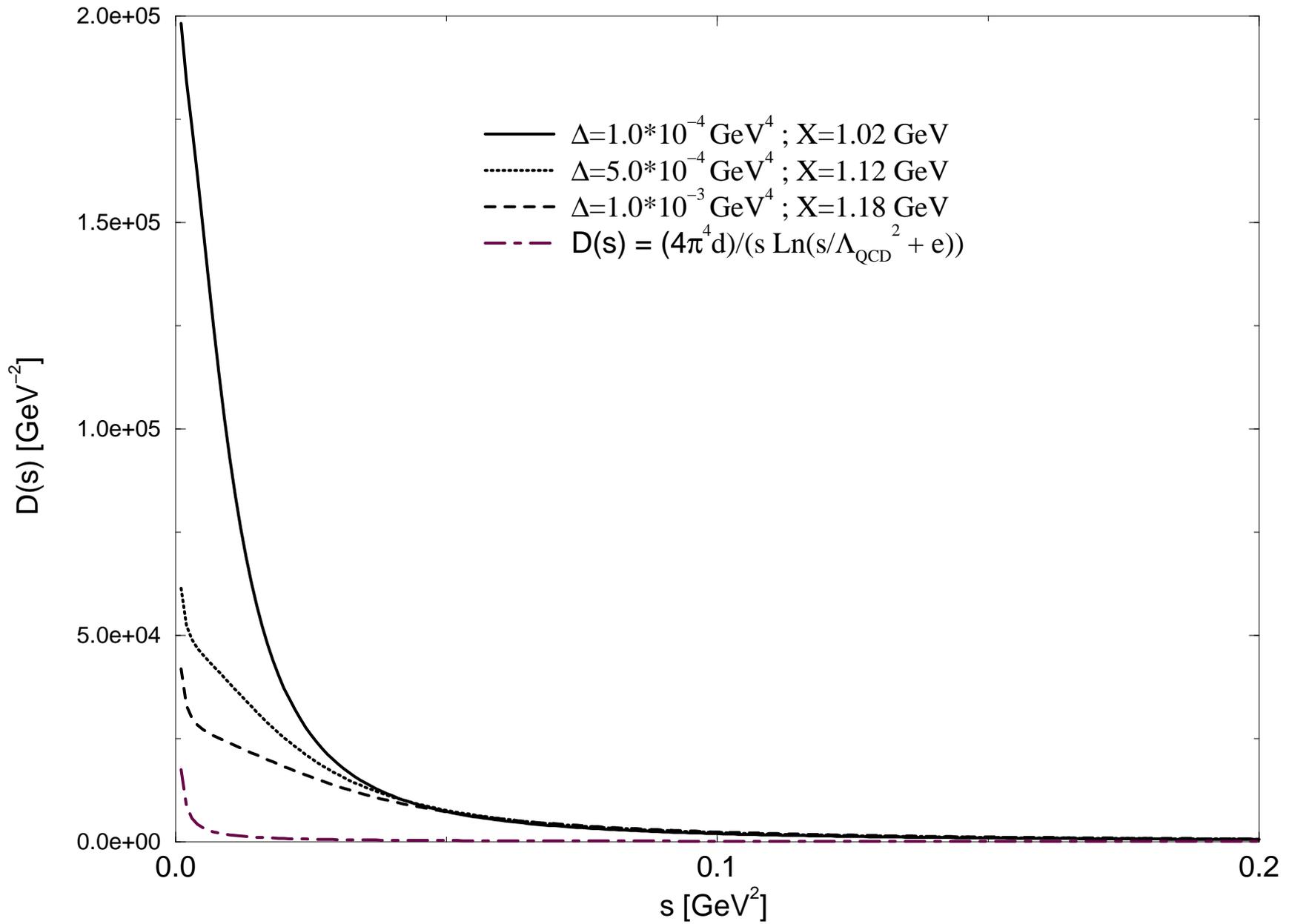
\begin{figure}
\caption{Model gluon-2 point functions 
$D(s)$
({\protect\ref{ddef}}) 
for three
the 3 different parameter sets 
({\protect\ref{sets}}).
In each case the value of the pion decay constant in the chiral limit 
$f_\pi = 88 \mbox{MeV}$ is reproduced. 
For comparison we also show the form of the UV tail 
$D_{UV} (s)$ ({\protect\ref{duv}})
(dashed-dotted).}
\label{fig4}
\end{figure}

Our model gluon propagator has the form
\begin{equation}
D(q^2) = D_{IR} (q^2) + D_{UV} (q^2) \, ,
\label{ddef}
\end{equation}
where
\begin{eqnarray}
{g_s}^2 D_{IR}(q^2) &=& (4\pi^2d) \frac{\chi^2}{q^4 + \Delta} \label{dir} \\
{g_s}^2 D_{UV}(q^2) &=& 
\frac{4\pi^2d}{q^2 + \mbox{Ln} \left ( \frac{q^2}{\Lambda_{QCD} ^2} + \tau 
\right )  } \; .
\label{duv}
\end{eqnarray}
The first term $D_{IR}$ (\ref{dir}), which dominates for small $q^2$,
is a regularized Mandelstam ansatz \cite{mandelstam,marciano}
with a strength $\chi^2$ and an IR regulator $\Delta$, which models
the IR strength of the quark-quark interaction. 
The second term $D_{UV}$ (\ref{duv}), which dominates for large $q^2$ 
is an asymptotic UV tail and matches the known 1-loop renormalization group 
result with $d=\frac{12}{33-2N_f} = \frac{12}{27}$, $\Lambda_{QCD} = 200
\mbox{MeV}$ and $\tau = e$ 
\cite{rw}.
The model parameters $\chi$ and $\Delta$ are adjusted to reproduce the weak 
pion decay constant in the chiral limit $f_\pi = 88 \mbox{MeV}$.
We are using 3 different parameter sets:
\begin{mathletters}
\label{sets}
\begin{eqnarray}
\mbox{Set 1:} \quad 
\Delta = 1.0*10^{-4} {\mbox{GeV}}^4 \; &;& \; \chi = 1.02 \mbox{GeV} 
\label{set1} \\
\mbox{Set 2:} \quad 
\Delta = 5.0*10^{-4} {\mbox{GeV}}^4 \; &;& \; \chi = 1.12 \mbox{GeV} 
\label{set2} \\
\mbox{Set 3:} \quad 
\Delta = 1.0*10^{-3} {\mbox{GeV}}^4 \; &;& \; \chi = 1.18 \mbox{GeV} \quad .
\label{set3} 
\end{eqnarray}
\end{mathletters}
The forms of $D(q^2)$ are displayed in Fig.\ref{fig4}, which also shows for 
comparison the pure UV form $D_{UV} (q^2)$. 
As one can see in all 3 cases 
the asymptotic form is reached already for $q^2 \approx 0.2 {\mbox{GeV}}^2$.
In Ref. \cite{fm1} it has been shown that with those values a satisfactorily 
description of all low energy chiral observables can be achieved.
 
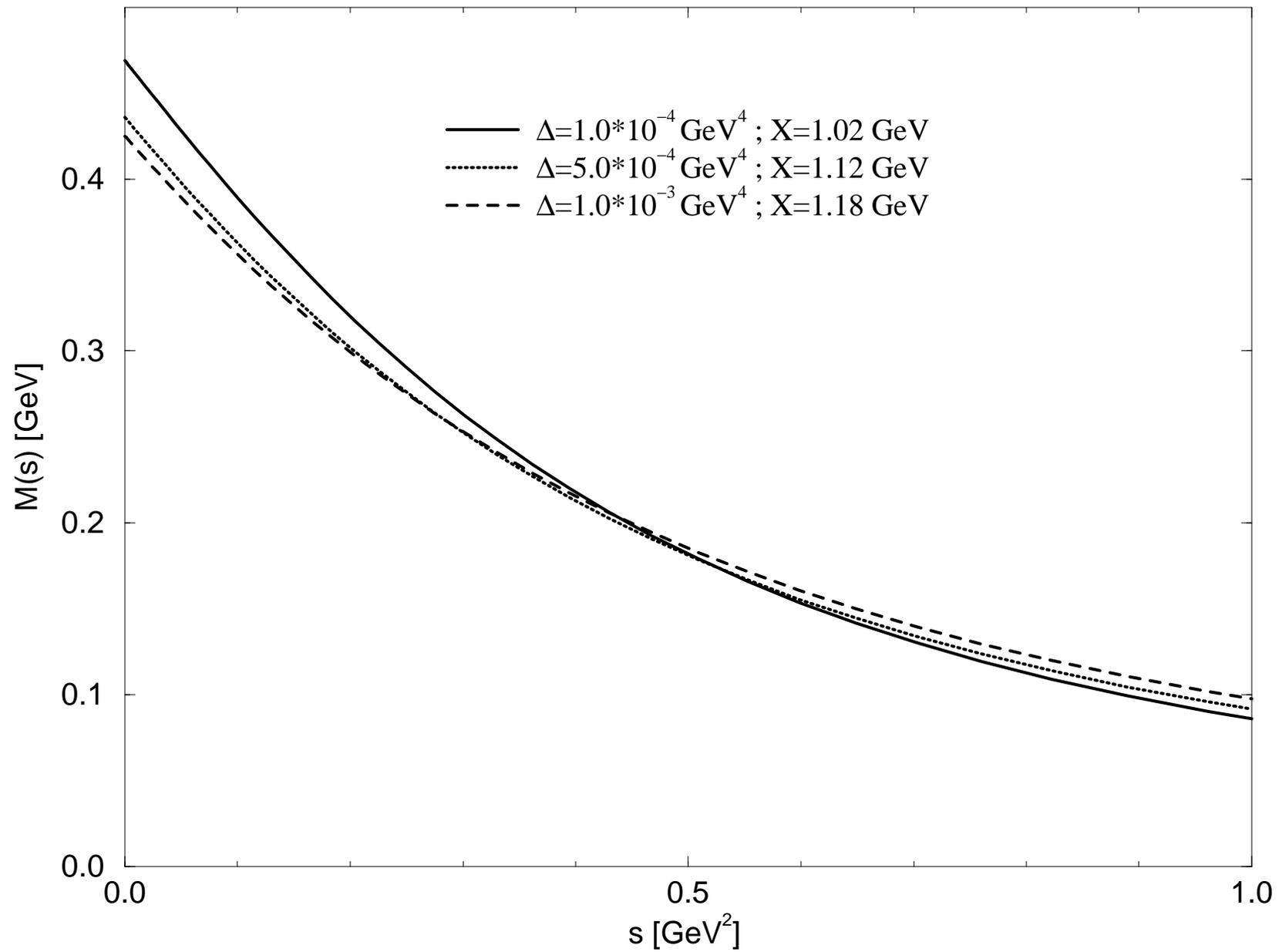
\begin{figure}
\caption{The dynamical (constituent) quark masses $M(s)$ for the three 
parameter sets   
({\protect\ref{sets}}).}
\label{fig5}
\end{figure}

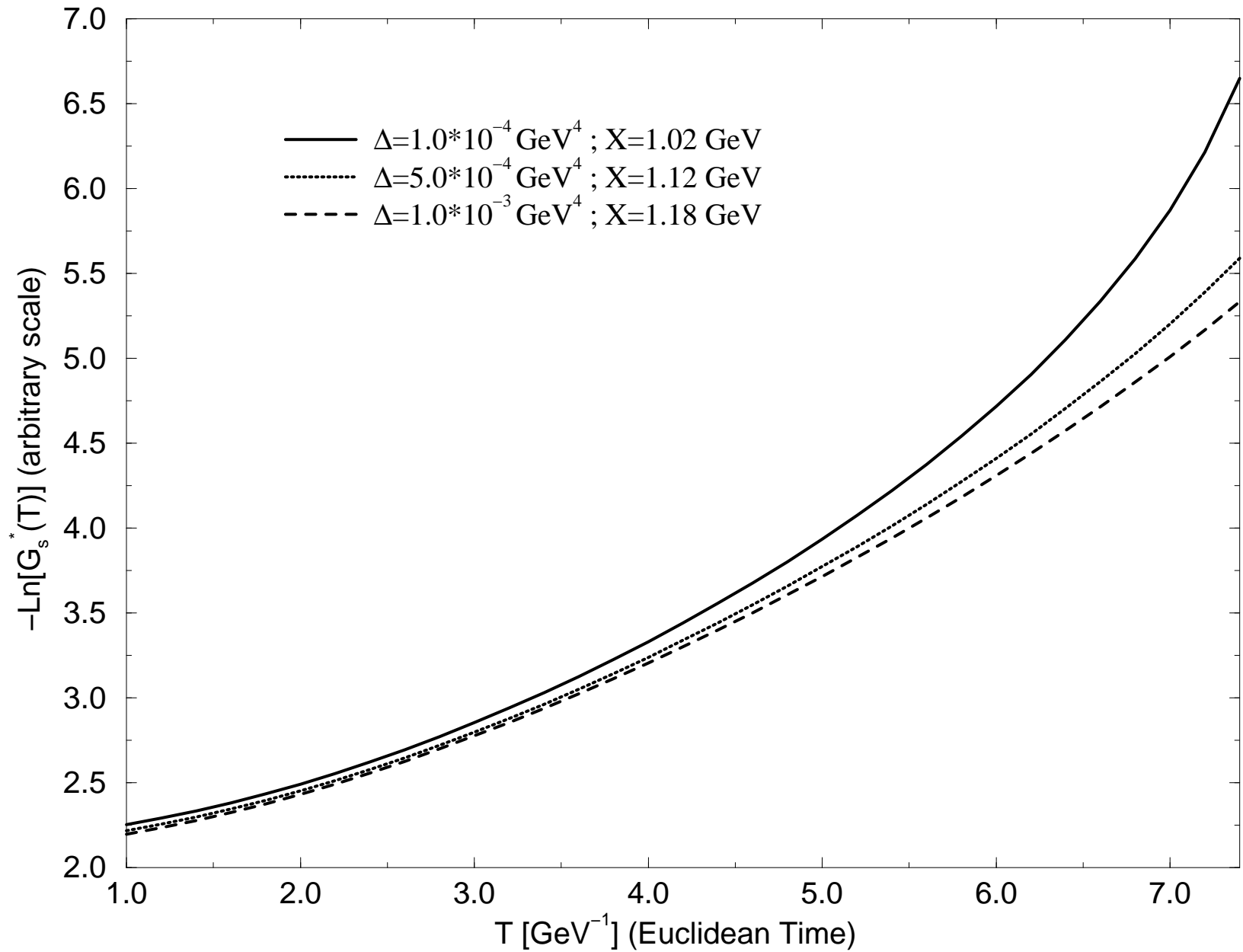
\begin{figure}
\caption{The Euclidean time Fourier transform ${G_S}^* (T)$ of the scalar part
of the quark propagator $G_s$ (arbitrary scale) 
for the three parameter sets   
({\protect\ref{sets}}).}
\label{fig6}
\end{figure}

The second step is the determination of the dressed quark propagator 
(\ref{quarkpropagator}) as solution of the coupled set of integral equations
(\ref{sde:a}) and (\ref{sde:b}).
Fig.\ref{fig5} displays the running constituent mass $M(p^2) = \frac{B(p^2)}{A
(p^2)}$ at spacelike $p^2>0$. In order to demonstrate that the quarks are 
confined one has to show
that the quark propagator $G(p^2)$ has no poles in the timelike region 
$p^2<0$. This can again be done by studying the Fourier transform with respect
to the Euclidean time $T$ \cite{rw}. It is sufficient to consider the scalar 
part $G_s$: 
\begin{equation} 
G_s ^* (T) = \int_{-\infty}^{+\infty} \frac{dq_4}{2\pi} e^{iq_4 T} G_s (q^2)  
= \int_{-\infty}^{+\infty} \frac{dq_4}{2\pi} e^{iq_4 T} 
\frac{B(q^2)}{q^2 A^2 (q^2) + B^2 (q^2)} \; .
\label{gtrans}
\end{equation}   
If $G(p)$ had a pole at $p^2 = -M^2$ the Fourier transform $G_s ^* (T)$ would 
fall off as 
$e^{-MT}$ for large $T$ or 
\begin{equation}
{\mbox{Ln}} [G_s ^* (T\to\infty)] \sim
- M T \label{gst}
\end{equation}
As we can see from Fig.\ref{fig6} $(-) \mbox{Ln} [G_s ^* (T)]$ is far from 
rising linearly with $T$,
at large $T$, it actually rises quadratically with $T$.
This shows that the the parameter set (\ref{sets}) produces indeed confined 
quarks.
 
The third step is now to solve the inhomogeneous BSE, 
i.e. the coupled set of 5 inhomogeneous linear integral equations 
(\ref{integraleq}).
We follow the numerical procedure from Ref. \cite{frank}.
One applies a Gauss quadrature using about $n_K\approx 30$ 
Gauss points for $\int_{-\infty}^{+\infty} dK$
and $n_{C_{Kq}}\approx n_{C_{KT}} \approx 10$  
Gauss points for both $\int_{-1}^{+1} dC_{Kq}$ and $\int_{-1}^{+1} dC_{KT}$.
It should be noticed that the angular integration over $C_{KT}$ is running
only over the the gluon propagator
$D$ and not over the functions
$\lambda_i (K^2,q^2,C_{KQ} ^2 )$. 
This transforms the integral equations into an high dimensional set of 
ordinary inhomogeneous linear equations for the $\lambda_i$ at the Gauss
points, which is solved by inverting the 
$(5\cdot n_K\cdot n_{C_{Kq}}) \times (5\cdot n_K\cdot n_{C_{Kq}}) 
\approx 1500 \times 1500$ 
coefficient matrix using standard packages \cite{lapack}.
 
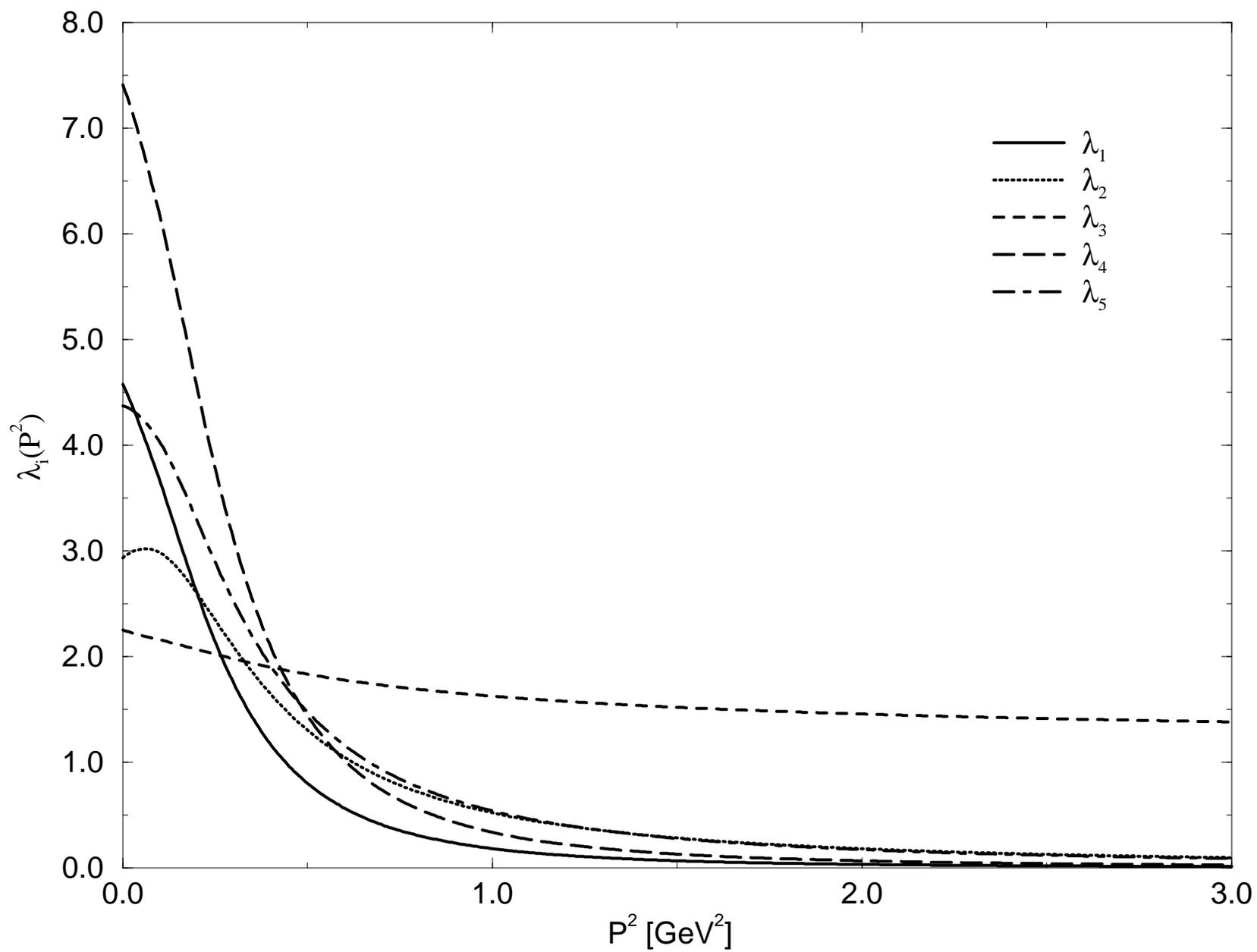
\begin{figure}
\caption{The five components $\lambda_i (P^2),\, i =1\dots 5$ of the vertex 
function $\Gamma_\mu (P,q)$ as defined in
({\protect\ref{decomp2}})
for a fixed spacelike value of the off shell momentum $q^2 = + 0.1 {\mbox{GeV}
}^2$ and angle
$C_{Pq} = 0$ using the parameter set 1 ({\protect\ref{set1}}).}
\label{fig7}
\end{figure}

Other than in Ref. \cite{frank}, where the quark self propagator $G$ 
which appears in the kernel of the BSE (\ref{ibsemomentum}) has been 
parameterized by a simple analytic form without solving
the DSE (\ref{sde}) explicitly, we are using the numerical solutions $A$ and 
$B$ of (\ref{sde}) in the kernel for the BSE (\ref{ibsemomentum}).
Our treatment is therefore a fully consistent solution 
of both the rainbow DSE for the quark propagator $G$
and the inhomogeneous ladder BSE for the vector vertex function $\Gamma_\mu$ 
using the same
model gluon propagator $D$ specified in (\ref{ddef}) as input.
As one can see from (\ref{ibsemomentum}) and (\ref{integraleq}) this involves 
a numerical interpolation of $A(P^2)$ and $B(P^2)$ to the points $P_{\pm}$.
Fig.\ref{fig7} displays the coefficient functions $\lambda_i (P^2)$ for the 
parameter set 1 (\ref{set1})
for a fixed spacelike value of $q^2 = 0.1 {\mbox{GeV}}^2$ and a fixed angle 
$C_{Pq} =0$.
As one can see our result is qualitatively similar though quantitatively 
noticeably different
than the one from Ref. \cite{frank}.
Similar than Ref. \cite{frank} we also find that the $\lambda_i(P^2,q^2,C_{PQ}
^2 )$ are only very weakly dependent on the angle $C_{PQ}$ if $P^2$ and
$q^2$ are fixed.
 
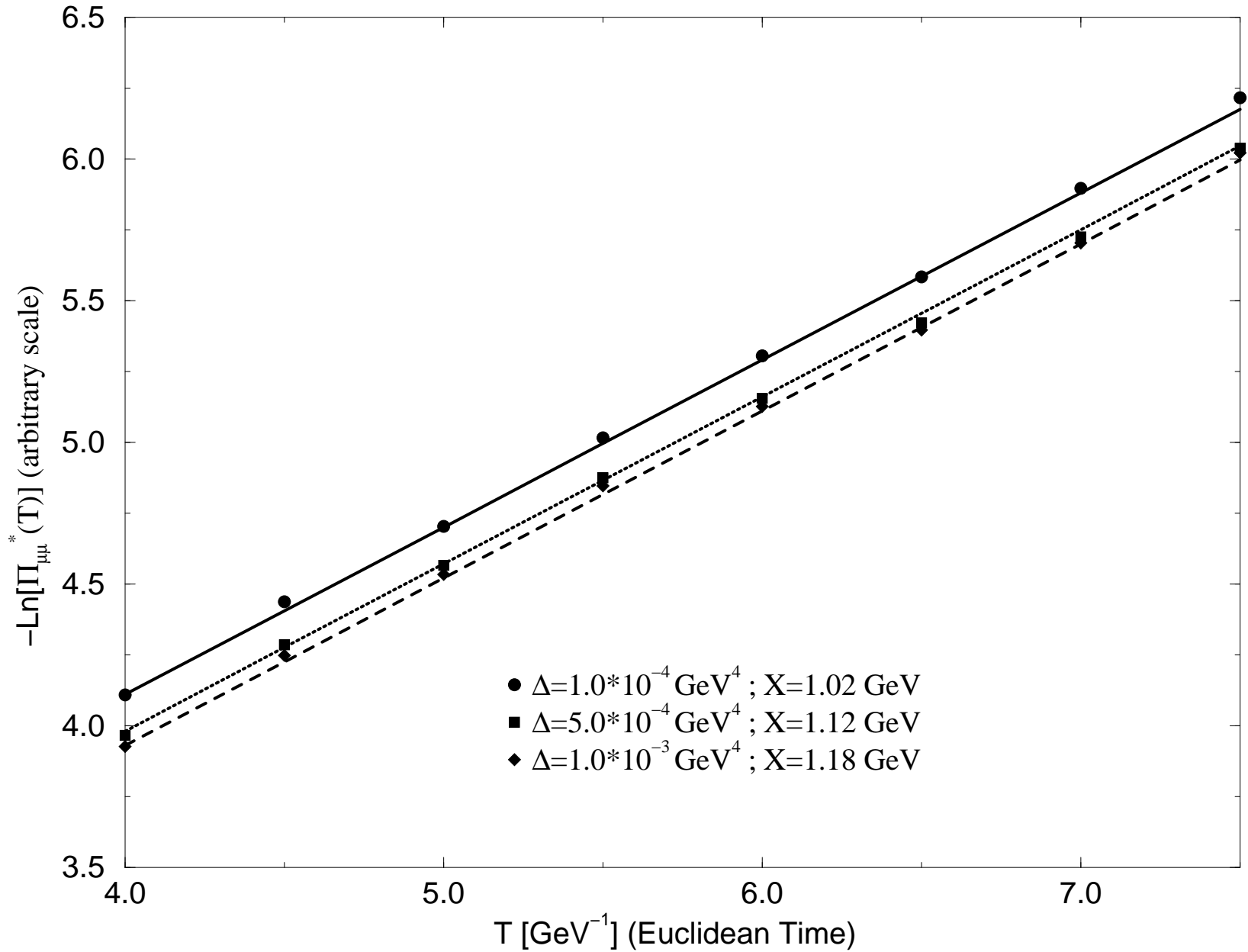
\begin{figure}
\caption{The Euclidean time Fourier transform ${\Pi_{\mu\mu}}^* (T)$ of the 
vector correlator (arbitrary scale) for the three parameter sets   
({\protect\ref{sets}}).
Our numerical results are denoted by circles, squares and diamonds, 
respectively.
The straight full, dotted and dashed straight lines are the best linear fits.}
\label{fig8}
\end{figure}

Finally Fig.\ref{fig8} shows our main result, the Fourier transformed 
correlator ${\Pi^*}_{\mu\mu} (T)$ for large Euclidean times $T$ evaluated
with the three sets (\ref{sets}).
As mentioned earlier it is essential to evaluate the Fourier integral 
$\int_{-\infty}^{+\infty} \frac{dq_4}{2\pi} e^{iq_4 T} $ before doing the 
integral over the quark momentum $\int_{-\infty}^{+\infty} dPP^3$ in order to
avoid UV infinities. From  Fig.\ref{fig8} we see clearly that ${\Pi^*}_{\mu\mu} (T)$ has an 
exponential falloff $\sim e^{-m_0 T}$for large $T$ and moreover that the
ground state mass $m_0$ is practically the same for all three
parameter sets, because the curves are parallel.
The ground state mass $m_0$ can be extracted by 
carrying out a linear fit to the numerical result.
One finds $m_0 \approx 590 \mbox{MeV}$ for all three sets.
This has to be compared with the phenomenological values of $780\mbox{MeV}$ 
for the ground 
state vector meson mass.
The instanton liquid model \cite{shuryak} finds
$m_0 = 950\mbox{MeV}$,
whereas quenched lattice QCD \cite{cghn} obtains $m_0 = 720\mbox{MeV}$.

\section{Conclusions}
To summarize: We have studied a $\bar{q}q$ bound state in the vector channel 
using a nonlocal and confining model quark-quark interaction, which respects 
all global symmetries of QCD.
The model parameters have been chosen to give a good description of low energy
chiral physics for
the Goldstone degrees of freedom ($\pi$, $K$, $\eta$).  
For the first time we have demonstrated the existence of a higher mass state 
by evaluating the correlator of 2 interpolating vector currents
in Euclidean space and showing the exponential falloff of its 
Fourier transform at large Euclidean times.
This employs a consistent treatment of the dressed quark propagator $G$ and 
the dressed vector vertex $\Gamma_\mu$,
which are both determined from the model quark-quark interaction by the 
rainbow Dyson-Schwinger equation for $G$
and the inhomogeneous ladder Bethe-Salpeter equation for $\Gamma_\mu$.
The method stays until the end in Euclidean space 
and all momenta are spacelike.
This avoids any unsafe analytic continuations
of the dressed propagators into the timelike region. 
We have found that the ground state mass is practically independent on the 
parameters of the model interaction and lies at about $590{\mbox{MeV}}$.
 
For the creation of a bound state it is obviously necessary to have an highly 
nonlocal quark-quark interaction,
which is reflected in its large nonperturbative strength in the IR. This 
leads not only to dynamical
chiral symmetry breaking but also to a strong momentum dependence of the
running dynamical quark mass $M(p^2)$ and, in our case even to confined
quarks. This is an important advantage over with the 
Nambu--Jona-Lasinio (NJL) interaction \cite{njl}, which can be regarded
as a special case of the GCM with a local quark-quark interaction 
$D_{\mu\nu} ^{ab} (x,y) = \delta^{ab} \delta_{\mu\nu} \delta^{(4)} (x-y)$ and 
has been subject to extensive studies \cite{njlreviews}.
The NJL model has a momentum independent constituent mass $M$ and is
therefore not confining.
Though it gives a satisfactorily description of the physics of associated 
with the Goldstone degrees of freedom  
($\pi$, $K$, $\eta$) it has trouble to deal with higher $\bar{q}q$ bound 
states, which are lying 
around the $2M$ threshold and are therefore unstable or barely stable against 
decay into a quark-antiquark pair.
 
As long as ground state properties are considered,
the method which we have used can also be applied to other channels, such as 
e.g. the axial vector mesons,
as well as to 3-point functions, which are the basis for the study of meson 
form factors \cite{bfm}.
The only implicit assumption which 
we have to make is the existence of 
a sufficiently large gap between
the ground state and the first excited state.
If ground state and first excited state are too close together, the 
exponential falloff is only
evident at very large numerical values of the Euclidean time $T$, where the 
result is likely
to be swamped by numerical errors in the Fourier transform due to the highly
oscillating behavior of $e^{iq_4 T}$.  
 
\acknowledgements
The authors would like to thank Matthias Burkardt (NMSU) 
and Mikkel Johnson (LANL) for helpful discussions and 
comments and 
Michael Frank for making the computer code used in Ref.\cite{frank} available.
 
This work was supported in part by the National Science Foundation grant
PHY-9319641.

\newpage
\appendix
\section{The Global Color Model (GCM) in an External Vector Field}
Following Ref. \cite{ft} we consider the Euclidean action of the Global Color Model (GCM) in an
{\it external vector field} ${\cal{A}}_\mu (x)$
\begin{equation}
S_{GCM} [{\bar{q}},q;{\cal{A}}]= 
\int d^4 x d^4 y \left \{
\bar{q} (x) [\gamma \cdot \partial_x - i \gamma_\nu {\cal{A}}_\nu (x) ] q (x) + \frac{{g_s}^2}{2} 
j_\mu ^a (x)
D_{\mu\nu}^{ab} (x-y) j_\mu ^b (y) \right \}
\label{sgcm}
\end{equation}
where $j_\mu^a $ denotes the color octet vector current:
\begin{equation}
j_\mu^a (x) = \bar{q} (x) \gamma_\mu \frac{\lambda_c^a}{2} q(x) \, .
\label{jdef}
\end{equation}
For convenience we will employ 
the Feynman-type gauge (\ref{feynmangauge}) \cite{rw,tandy} 
for the model gluon propagator $D$.
Applying the standard bosonization procedure 
\cite{kleinert,shrauner} the generating functional 
\begin{equation}
{\cal{Z}} [{\cal{A}} ]  \equiv e^{- {\cal{W}} [{\cal{A}} ] }  = 
\int  
{\cal{D}} {\bar{q}} {\cal{D}} q
e^{- S_{GCM} [{\bar{q}},q;{\cal{A}}] } 
\label{zgcm1}
\end{equation}
can be rewritten in terms of the bilocal auxiliary fields ${\cal{B^\theta}}(x,y)$
\begin{equation}
{\cal{Z}} [{\cal{A}} ] = \int {\cal{D}} {\cal{B^\theta}} e^{- S_{eff} [{\cal{B^\theta}};{\cal{A}} ] }
\label{zgcm2}
\end{equation}
with the effective bosonic action
\begin{equation}
S_{eff} [{\cal{B^\theta}};{\cal{A}} ] = (-) \mbox{Tr} \mbox{Ln} {\cal{G}}^{-1} [{\cal{B^\theta}};{\cal{A}} ]
+ \int d^4 x d^4 y \frac{ {\cal{B^\theta}}(x,y)  {\cal{B^\theta}}(y,x) }{2{g_s}^2 D(x-y) }  
\label{seff}
\end{equation}
and the quark operator
\begin{equation}
{\cal{G}}^{-1} [{\cal{B^\theta}};{\cal{A}} ] = 
\left [ \gamma \cdot \partial_x - i \gamma_\nu {\cal{A}}_\nu (x) \right ] \delta (x-y) + \Lambda^\theta  
{\cal{B^\theta}} (x,y) \, .
\label{gdef}
\end{equation} 
The matrices $\Lambda^\theta$ arise from Fierz reordering the current-current interaction in 
(\ref{sgcm}) and are given by  
\begin{equation}
\Lambda ^{\theta }=\frac{1}{2}\left( {\bf 1}_{D},
i\gamma 
_{5},\frac{i}{\sqrt{2}}\gamma _{\nu },\frac{i}{\sqrt{2}}
\gamma _{\nu}\gamma _{5}\right) \otimes \left( \frac{1}{\sqrt{3}}{\bf 1}_{F},
\frac{1}{\sqrt{2}}\lambda _F^a\right) \otimes \left( \frac{4}{3}{\bf 1}_{C},
\frac{i}{\sqrt{3}}\lambda _c^{a}\right)
 \, .
\label{fierz}
\end{equation}  
In the mean field approximation, 
which is the leading order in $\frac{1}{N_c}$, the fields ${\cal{B^\theta}} (x,y)$
are substituted by their vacuum values  
${\cal{B}}_0 ^\theta(x,y)$ which are given as the stationary points of  
the effective action (\ref{seff}) 
\begin{equation}
\left [ \frac{\delta S_{eff}}{\delta {\cal B}} \right ]_{{\cal B}_0}=0
\label{stationary1}
\end{equation}
or
\begin{equation}
{\cal{B}}_0 ^\theta[{\cal{A}}] (x,y) = {g_s}^2 D(x-y) \mbox{tr} [
\Lambda^\theta 
{\cal{G}}_0 [{\cal{A}}] ^{-1} (x,y)
]  \, ,
\label{stationary2}
\end{equation}
where ${\cal{G}}_0 ^{-1} (x,y)$ denotes the inverse propagator 
with the self energy $\Sigma(x,y) = \Lambda^\theta {\cal{B}}_0 ^\theta(x,y)$
in the external background field ${\cal{A}} (x)$
\begin{equation}
{\cal{G}}_0 [{\cal{A}}] ^{-1} (x,y) = 
\left [ \gamma \cdot \partial_x - i \gamma_\nu {\cal{A}}_\nu (x) \right ] \delta (x-y) + \Lambda^\theta  
{\cal{B}}_0 ^\theta(x,y) \, .
\label{g0def}
\end{equation}
We want to stress that both ${\cal{B}}_0 ^\theta(x,y)$ as well as ${\cal{G}}_0 ^{-1} (x,y)$
have an implicit dependence on the external background field ${\cal{A}} (x)$.
If the external field ${\cal{A}}$ is switched off  ${\cal{G}}_0$ goes into the dressed
quark propagator $G \equiv {\cal{G}}_0 [ {\cal{A}}=0] $ which has the decomposition (\ref{quarkpropagator}):
\begin{equation}
G^{-1} (p) \equiv   i \gamma \cdot p [A(p^2) -1] + \Sigma(p) 
\equiv i \gamma \cdot p A(p^2) + B(p^2) 
\end{equation}
in momentum space.
The quark self energy dressing $\Sigma (p)$ 
is determined as the solution of the rainbow DSE (\ref{stationary2} for
${\cal{A}}=0$) rendering (\ref{sdesigma}):
\begin{equation}
\Sigma (p) = \frac{4}{3}{g_s}^{2} \frac{d^{4}q}{(2\pi)^{4}}
D (p-q) \gamma_\nu G (q) \gamma_\nu \, .
\end{equation}

\section{Vertex Dressing and Inhomogeneous Bethe-Salpeter Equation}
In coordinate space the dressed vector vertex $\Gamma_\mu (y_1,y_2;z)$ is given as
the functional derivative of the inverse quark propagator ${\cal{G}}_0 [{\cal{A}}] ^{-1}$
(\ref{g0def})
with respect to the external field ${\cal{A}}_\mu$: 
\begin{equation} 
\Gamma_\mu (y_1,y_2;z) \equiv 
\left [
\frac{\delta {\cal{G}}_0 [{\cal{A}}] ^{-1} (y_1,y_2) }{\delta {\cal{A}}_\mu (z) }
\right ]_{ {\cal{A}} =0} \,  .
\label{gammadef}
\end{equation}
Taking the functional derivative in (\ref{g0def}) gives for
(\ref{gammadef}):
\begin{equation}
\Gamma_\mu (y_1,y_2;z) = 
(-i) \gamma_\mu \delta(y_1 - y_2) \delta(y_1 - z) + \left [
\frac{\delta \Sigma [ {\cal{A}} ] (y_1, y_2) }{\delta{\cal{A}} (z) } \right ]_{ {\cal{A}} =0 } \, .  
\label{ibseder1}
\end{equation}
The second term on the r.h.s. of (\ref{ibseder1}) can be determined 
by employing
the stationary condition (\ref{stationary2}), which 
after Fierz reordering can be cast into
\begin{equation}
\Sigma[{\cal{A}}] (y_1,y_2) = \frac{4}{3} {g_s}^2 D(y_1-y_2) \gamma_\nu
\left [
\frac{\delta {\cal{G}}_0 [{\cal{A}}]  (y_1,y_2) }{\delta {\cal{A}}_\mu (z) }
\right ]_{ {\cal{A}} =0} 
\gamma_\nu \, .
\label{ibseder2}
\end{equation}
In order to find an expression for 
$\left [
\frac{\delta {\cal{G}}_0 [{\cal{A}}]  (y_1,y_2) }{\delta {\cal{A}}_\mu (z) }
\right ]_{ {\cal{A}} =0}$ in terms of the quark propagator $G$.
we write the definition (\ref{g0def}) schematically as
\begin{equation}
{\cal{G}}_0 ^{-1} [{\cal{A}}] = G^{-1} + {\cal{A}}_\mu \Gamma_\mu
\label{ibseder4}
\end{equation}  
which leads to the formal expansion
\begin{equation}
 {\cal{G}}_0 [{\cal{A}}]  = G - G {\cal{A}}_\mu \Gamma_\mu G \pm \dots \, .
\label{ibseder3}
\end{equation}  
Putting (\ref{ibseder3}) into (\ref{ibseder2}) gives
\begin{equation}
\left [
\frac{\delta \Sigma [ {\cal{A}} ] (y_1, y_2) }{\delta{\cal{A}} (z) } \right ]_{ {\cal{A}} =0 }   
=
- 
\frac{4}{3} {g_s}^2 D(y_1 - y_2) \int d u_1 d u_2 
\gamma_\nu  G(y_1,u_1)
\Gamma_\mu (u_1,u_2;z) G(u_2,y_2) \gamma_\nu \, ,
\label{ibseder5}
\end{equation}
which, after substituting the result into (\ref{ibseder1}) renders the
{\it inhomogeneous} BSE for  $\Gamma_\mu (y_1,y_2;z)$ 
in coordinate space
\begin{eqnarray}
&& 
\Gamma_\mu (y_1,y_2;z) \, = \, \\ \nonumber
&&
(-i) \gamma_\mu \delta(y_1 - y_2) \delta(y_1 - z) - 
\frac{4}{3} {g_s}^2 D(y_1 - y_2) \int d u_1 d u_2  
\gamma_\nu  G(y_1,u_1)
\Gamma_\mu (u_1,u_2;z) G(u_2,y_2) \gamma_\nu \, .
\label{ibsecoordinate}
\end{eqnarray}
Fourier transform leads then to the momentum space form (\ref{ibsemomentum}).

\section{Matrix Elements of the BS Kernel}
In this appendix we list the explicit forms for the 
quantities which appear as kernel  
of the 5 transversal inhomogeneous BS integral
and will be also needed in (\ref{picomp}):  
\begin{equation}
{\cal{X}} = (0,3,1,0,0) \,
\label{xdef}
\end{equation}
\begin{equation}
\begin{array}{ll}
{\cal N}_{11}=1&
{\cal N}_{22}=-\frac{(P^T)^2}{\eta ^2}\\
{\cal N}_{23}=3&
{\cal N}_{32}=-\frac{(P^T)^2}{\eta ^2}\\
{\cal N}_{33}=1&
{\cal N}_{42}=1\\
{\cal N}_{44}=\frac{q^2}{\eta ^2}&
{\cal N}_{55}=1   
\end{array}
\label{ndef}
\end{equation}
\nonumber \\
and
\begin{equation}
\begin{array}{ll}
{\cal M}_{11}=2\pi \hat{D}_1F_0\frac{K}{P^T}&
{\cal M}_{12}=2\pi \hat{D}_1\frac{K^3}{P^T\eta
^2}\left(V+\frac{q^2}{\eta ^2}C_{Kq}^2W\right) \\
{\cal M}_{13}=-2\pi \hat{D}_1\frac{K}{P^T}V&
{\cal M}_{14}=2\pi \hat{D}_1\frac{K^3q^^2}{P^T\eta
^4}C_{Kq}^2\left(V+\frac{q^2}{\eta ^2}W\right) \\
{\cal M}_{21}=-\pi \hat{D}_0\frac{K^2}{\eta ^2}x^2V&
{\cal M}_{22}=\pi \hat{D}_0\frac{K^2}{\eta ^2}x^2\left(
F_1-2\frac{K^2}{\eta ^2}T\right) \\
{\cal M}_{23}=-\pi \hat{D}_0\left( 3F_1-2\frac{K^2}{\eta ^2}T\right) &
{\cal M}_{24}=-2\pi \hat{D}_0\frac{K^4q^2}{\eta ^6}x^2C_{Kq}^2T \\
{\cal M}_{25}=2\pi \hat{D}_0\frac{K^2q^2}{\eta ^4}x^2T &
{\cal M}_{31}=-\pi \hat{D}_2\frac{K^2}{\eta ^2}V \\
{\cal M}_{32}=\pi \hat{D}_2\frac{K^2}{\eta ^2}\left(
F_1-2\frac{K^2}{\eta ^2}T\right) &
{\cal M}_{33}=-\pi \left( F_1\hat{D}_0-2\frac{K^2}{\eta ^2}T\hat{D}_2\right) 
\\
{\cal M}_{34}=-2\pi \hat{D}_2\frac{K^4q^2}{\eta ^6}C_{Kq}^2T&
{\cal M}_{35}=\pi \left( \hat{D}_0x^2-\hat{D}_2\right) \frac{K^2q^2}{\eta
^4}T\\
{\cal M}_{41}=\pi \hat{D}_1\frac{K^2}{P^LP^T}C_{Kq}\left(
V+\frac{q^2}{\eta ^2}W\right) &
{\cal M}_{42}=-\pi \hat{D}_1\frac{K^2}{P^LP^T}C_{Kq}F_2\\
{\cal M}_{43}=-2\pi \hat{D}_1\frac{K^2}{P^LP^T}C_{Kq}T&
{\cal M}_{44}=-\pi \hat{D}_1\frac{K^2}{P^LP^T}C_{Kq}\frac{q^2}{\eta ^2}F_3\\
{\cal M}_{53}=-\pi \hat{D}_1\frac{K}{P^T}T&
{\cal M}_{55}=\pi \hat{D}_1\frac{K}{P^T}F_0
\end{array}
\end{equation}
We have used the abbreviations:
\begin{equation}
\begin{array}{ll}
F_0\equiv \left( \frac{q^2}{4}-K^2\right) \frac{1}{\eta ^2}T+U&
F_1\equiv \left( K^2-\frac{q^2}{4}\right) \frac{1}{\eta ^2}T+U\\
F_2\equiv F_1+2\left( \frac{q^2}{4}-K^2\right) \frac{1}{\eta ^2}&
F_3\equiv F_1+2\left( \frac{q^2}{4}-K^2C_{Kq}^2\right) \frac{1}{\eta ^2}\\
T\equiv \eta ^4\alpha (K^2_{+})\alpha (K^2_{-})&
U\equiv \eta ^2\beta (K^2_{+})\beta (K^2_{-})\\
V\equiv \eta ^3\left[ \alpha (K_{+})\beta (K_{-})+\alpha (K_{-})\beta
(K_{+})\right] &
W\equiv \eta ^5 \left[ \frac{\alpha (K_{+})\beta (K_{-})-\alpha
(K_{-})\beta (K_{+})}{2K\cdot q}\right] 
\end{array}
\end{equation}
as well as
\begin{equation}
\alpha(p^2) = \frac{A(p^2)}{p^2 A^2(p^2) + B^2 (p^2) } \; , \; 
\beta(p^2) = \frac{B(p^2)}{p^2 A^2(p^2) + B^2 (p^2) } \; , \;
\end{equation}
and
\begin{eqnarray}
&& \hat{D}_n( P^2,C_{Pq},K^2,C_{Kq})
\equiv \nonumber \\
&& 
\int _{-x}^{x}
dC_{KT}
\frac{1}{3\eta^2\pi^4} \,
{g_s}^2
D \left( P^2 + K^2 - 2PK
\sqrt{1-C^2_{Pq}}C_{KT}- 2PKC_{Pq}C_{Kq}\right) (C_{KT})^n  \; .
\end{eqnarray}
The $C_{KT}$ is the direction cosine between $K_\mu$ and $P_\mu ^T$ and 
\begin{equation}
x \equiv \sqrt{1-C_{Kq} ^2 } \; , \; P_L \equiv P C_{PQ} \; , \; P_T = P \sqrt
{1-C_{Kq} ^2 } \; .
\end{equation}

\section{}  
In this appendix we give a derivation of Eq.(\ref{corr3}).
The time ordered product of the two vector currents can be formally written as
functional derivative of the generating functional 
${\cal{Z}} [{\cal{A}} ]$ (\ref{zgcm1},\ref{zgcm2}) 
with respect to the external vector
field ${\cal{A}}$:
\begin{equation}
\left \langle  {\cal{T}} j_{\mu_1} (z_1) j_{\mu_2} (z_2) \right \rangle =
\frac{1}{{\cal{Z}}[0]} (-) 
\left [\frac{\delta^{(2)} 
{\cal{Z}} [{\cal{A}} ]   }
{ \delta {\cal{A}}_{\mu_1}(z_1) \delta {\cal{A}}_{\mu_2}(z_2) }   
\right ]_{ {\cal{A}}=0}  \, .
\end{equation}  
At the mean field level the integration in 
over all possible configurations of the bosonic
auxiliary field ${\cal{B}}$ is substituted by the stationary configuration  
${\cal{B}}_0$ and we can therefore write:
\begin{equation}
{\cal{Z}} [{\cal{A}} ] = e^{-{\cal{W}} [{\cal{A}} ] } \approx  e^{-{\cal{W}}_0
 [{\cal{A}} ] } \; ,
\end{equation}
where
\begin{equation}
{\cal{W}}_0 [{\cal{A}} ] = 
(-) \mbox{Tr} \mbox{Ln} 
{\cal{G}}_0 [{\cal{A}}] ^{-1} 
+ \int d^4 x d^4 y \frac{ {\cal{B}}_0 ^\theta (x,y)  {\cal{B}}_0 ^\theta (y,x)
 }{2{g_s}^2 D(x-y) }  \, .
\end{equation}
This implies for the functional derivatives
\begin{eqnarray}
\frac{\delta {\cal{Z}} [{\cal{A}} ] }{\delta{\cal{A}}_\nu (z) }
&=&
(-) \frac{\delta {\cal{W}}_0 [{\cal{A}} ] }{\delta{\cal{A}}_\nu (z) }
e^{-{\cal{W}}_0 [{\cal{A}} ] } \nonumber \\
\left [ 
\frac
{\delta {\cal{W}}_0 [{\cal{A}} ] }
{\delta{\cal{A}}_\nu (z) }
\right ] &=& 
\left [ 
\frac{\partial {\cal{W}}_0 [{\cal{A}} ] }{\partial{\cal{A}}_\nu (z) }
\right ]
+ 
\left [ 
\frac{\delta {\cal{W}}_0 [{\cal{A}} ] }{\delta{\cal{B}}_0 [{\cal{A}}] } 
\right ]
\left [ \frac{ \delta{\cal{B}}_0 [{\cal{A}}] }{\delta{\cal{A}}_\nu (z) } 
\right ]
\nonumber \\
&=&
\left [ 
\frac
{\partial {\cal{W}}_0 [{\cal{A}} ] }
{\partial{\cal{A}}_\nu (z) }
\right ]
\nonumber \\
&=&
(+i) \langle z \vert \mbox{tr}
\left [ {\cal{G}}_0 [ {\cal{A}} ] \gamma_\nu \right ] \vert z \rangle 
\nonumber \\ 
\left [ 
\frac
{\delta {\cal{W}}_0 [{\cal{A}} ] }
{\delta{\cal{A}}_\nu (z) }
\right ]_{ {\cal{A}}=0} &=& 
(+i) \mbox{tr} \left [ G(z,z) \gamma_\nu \right ]  = 0 
\nonumber \\
\left [ \frac{\delta {\cal{Z}} [{\cal{A}} ] }{\delta{\cal{A}}_\nu (z) }
\right ]_{ {\cal{A}}=0} &=& 0  \nonumber \\
\left [
\frac{\delta^{(2)} 
{\cal{Z}} [ {\cal{A}} ]   }
{ \delta {\cal{A}}_{\mu_1}(z_1) \delta {\cal{A}}_{\mu_2}(z_2) }   
\right ]_{ {\cal{A}}=0}
&=& (-)  {\cal{Z}} [0] \;
\left [
\frac
{\delta^{(2)} 
{\cal{W}}_0 [{\cal{A}} ]   }
{ \delta {\cal{A}}_{\mu_1}(z_1) \delta {\cal{A}}_{\mu_2}(z_2) }   
\right ]_{ {\cal{A}}=0} \; .
\end{eqnarray}
Therefore we find for the correlator 
using (\ref{gammadef}) 
\begin{equation}
\left \langle  {\cal{T}} j_{\mu_1} (z_1) j_{\mu_2} (z_2) \right \rangle = i N_c  \,
\int d^4 y_1 d^4 y_2 {\mbox{tr}}_\gamma \left [
\gamma_{\mu_1} G(z_1,y_1) \Gamma_{\mu_2} (y_1,y_2;z_2) G(y_2,z_1) \right ] \, ,
\end{equation}
which, after Fourier transform, gives (\ref{corr3}).

\end{document}